\documentclass[prd,12pt,preprintnumbers,nofootinbib,showpacs]{revtex4}
\usepackage{epsfig,amsmath,amssymb}
\def\KeyWord#1{$\backslash$\IfColor{$\!\!$\textRed{#1}\textBlack}{#1}$\!\!$}
\usepackage{epsfig,amsmath}

\def\Q1{{\bf Q}_1}

\begin{document}
\preprint{\bf FIRENZE-DFF - 411/02/2004}

\title{A calculation of the QCD phase diagram at finite temperature, and baryon
and isospin chemical potentials}
\author{A. Barducci, R. Casalbuoni, G. Pettini, L. Ravagli}

\affiliation{Department of Physics, University of Florence and
INFN
Sezione di Firenze \\
Via G. Sansone 1, I-50109, Sesto F.no, Firenze, Italy}
\email{barducci@fi.infn.it, casalbuoni@fi.infn.it,
pettini@fi.infn.it, ravagli@fi.infn.it}

\begin{abstract}                
We study the phases of a two-flavor Nambu-Jona-Lasinio model at
finite temperature $T$, baryon and isospin chemical potentials:
$\mu_{B}=(\mu_{u}+\mu_{d})/2~~$, $\mu_{I}=(\mu_{u}-\mu_{d})/2$.
This study completes a previous analysis where only small isospin
chemical potentials $\mu_{I}$ were considered.
\end{abstract}
  \maketitle

\section{INTRODUCTION}
\label{int}
The possible formation of a pion condensate due to a finite
isospin chemical potential $\mu_{I}$ has been, in recent years,
the subject of several papers
\cite{Buballa:1998pr,Bedaque:1999nu,Toublan:1999hx,Steiner:2000bi,Son:2000xc,Alford:2000ze,
Splittorff:2000mm,toub2001,Steiner:2002gx,Neumann:2002jm,
Klein:2003fy,Toublan:2003tt,Frank:2003ve,Barducci:2003un}.
Consequently, the reconstruction of the QCD phase diagram at
finite temperature and quark densities, such as those attainable
in earth experiments and in the interior of stars, is even more
challenging due to the addition of $\mu_{I}$, besides temperature
$T$ and baryon chemical potential $\mu_{B}$. Various regions of
the phase diagram correspond to different experimental settings.
Actually, the behavior of QCD at high temperature and low baryon
densities is central to the relativistic heavy-ion collisions:
experiments at CERN and RHIC are expected to produce hadronic
matter in this regime. On the other hand, the description of
neutron star interiors requires the knowledge of cold nuclear
matter  at large baryon densities. However, nature does also
provide us with systems at finite isospin chemical potential
$\mu_{I}$ in the form of asymmetric-isospin matter inside neutron
stars: nuclear matter has a finite (negative) isospin $I_{3}$
density due to Coulomb interactions, apart from finite
baryon-number density. Moreover, in any realistic experimental
setting in relativistic heavy-ion collisions there is a non-zero,
even if small, $\mu_{I}$.\\
Our present description of the QCD phase diagram in the plane
$(\mu_{B},T)$ anticipates the existence of a tricritical point
separating first order transitions in the regions of low
temperatures from cross-over transitions in the low baryon
chemical potential and high temperature regime
\cite{Barducci:1989wi,Barducci:1990wi,klevansky,Barducci:1993bh,Kunihiro,Halasz:1998qr,Berges:1998rc}.
In recent years various other non-trivial phases of QCD at low
temperatures and high baryon chemical potentials have been
discovered, such as the Color Flavor Locking (CFL) phase and the
two-color superconducting (2SC) phase (for a review see
\cite{Rajagopal:2000wf}).

Coming back to the effects of the isospin chemical potential
$\mu_{I}$, pion condensation has so far been primarily
investigated by means of low energy models based on chiral
lagrangians
\cite{Splittorff:2000mm,Son:2000xc,toub2001,Jakovac:2003ar}.
Although these models are well suited to study the phases of QCD
as they have the right symmetry properties, they do not include
the combined effects of the isospin chemical potential $\mu_{I}$
with a finite baryon chemical potential $\mu_{B}$ in order to
study the pattern of chiral symmetry breaking and restoration as
well. To consider both $\mu_{B}$ and $\mu_{I}$, we need a model
with quarks as microscopic degrees of freedom. The effect of small
$\mu_{I}$ (up to half of the pion mass) has been investigated in
ref. \cite{Toublan:2003tt} in the context of the
Nambu-Jona-Lasinio model and in {\it ladder}-QCD
\cite{Barducci:2003un}. The result is the splitting of the
critical curves for chiral symmetry restoration for the two light
flavors, whereas a full study for arbitrary $\mu_{I}$ has only
been done in the context of a random matrix model
\cite{Klein:2003fy}.

Studies on the lattice have been performed at finite $\mu_{I}$ and
$\mu_{B}=0$ in refs.
\cite{Kogut:2002tm,Kogut:2002zg,gupta,deforcrand,Sinclair:2003rm}
and with a finite $\mu_B$ and $\mu_I=0$ in refs.
\cite{Fodor:2001au,Fodor:2001pe,allton,D'Elia:2002gd,Sinclair:2003rm}.
In a recent work \cite{nishida} the effect of both $\mu_{B}$ and a
small $\mu_{I}$ has also been considered. The case of high $\mu_B$
and small $\mu_I$ has been considered in \cite{Hands:2004uv}.

In this work we extend the analysis of
\cite{Toublan:2003tt,Barducci:2003un} where it was found that the
first order transition line ending at the tricritical point of the
case $\mu_I=0$ actually splits into two first order transition
lines and correspondingly two crossover regions are present at low
values of baryon chemical potential. In particular we will be
working in the context of a NJL model with a form factor included
such as to imply a decreasing of the fermion self-energy
compatible with the operator product expansion.

 It should also be noticed
that in \cite{Frank:2003ve} the NJL model has been augmented by
the four-fermi instanton interaction relevant in the case of two
flavors. These authors have found that the coupling induced by the
instanton interaction between the two flavors might completely
wash the splitting of the first order transition line. This
happens for values of the ratio of the instanton coupling to the
NJL coupling of order 0.1-0.15.

In Section II we summarize the relevant features of the NJL model
we have considered, with isospin charge included. The one-loop
effective potential and the values of the fit parameters are
included. In Section III we discuss the various equilibrium phases
of the model, together with the corresponding symmetries, by
studying the behavior of the scalar and pion condensates with
respect to different thermodynamical parameters among
$T,\mu_B,\mu_I$ (or $\mu_u,\mu_d$). Results are shown for growing
temperatures, starting from zero up to temperatures above that of
the critical ending point. Finally, Section IV is devoted to
conclusions.

\section{The model}
\label{sec:physics}

Our purpose is to explore the structure of the phase diagram for
chiral symmetry and pion condensation in QCD at finite temperature
and quark densities, by using a microscopic model with quark
degrees of freedom. This task has been accomplished, up to now, in
the context of a random matrix model simulating QCD with two
flavors \cite{Klein:2003fy} and, in the case of small differences
between the $u$ and $d$ quark chemical potentials, also in the
Nambu-Jona-Lasinio model (NJL) \cite{Toublan:2003tt} and in
$ladder$-QCD
\cite{Barducci:2003un}.\\
One reason for using a model with quarks as microscopic degrees of
freedom is that it gives us the possibility of studying chiral
symmetry breaking and pion condensation at both finite isospin and
baryon chemical potentials, which is not possible within effective
chiral models.\\ In ref. \cite{Toublan:2003tt}, the authors made
use of the NJL model with a suitable form factor included in the
quark self-energy to mimic asymptotic freedom
\cite{Alford:1997,Berges:1998rc}. This version of the NJL model
turns out to be very close to $ladder$-QCD as developed in refs.
\cite{Barducci:1988gn,Barducci:1990wi} where the momentum
dependence of the quark self-energy is consistently dictated by
the study of the Schwinger-Dyson equation within a variational
approach (see the previous references for details). However,
although $ladder$-QCD is a covariant and self-consistent approach,
the dependence on the four-momentum of the quark self-energy makes
the numerical computation of the one-loop effective potential with
finite quark densities much more onerous with respect to the NJL
case, where the quark self-energies depend only on the
three-momentum. For this reason, in the present work we study the
NJL model. It is reasonable to expect that when employing
$ladder$-QCD, the resulting physical picture does not considerably
differ from that of the NJL model. This has been the case in
previous applications too
\cite{Barducci:1990wi,klevansky,Kunihiro}.

As already said in the Introduction we are not going to consider
the effects of di-fermion condensation. Therefore our results can
be considered valid only outside the region of the color
superconductive phase, which is roughly in the region defined by
$\mu_{B}\gtrsim 400-500~MeV$ and $T\lesssim 50~MeV$. At the same
time we will not consider regions at values of $\mu_{u}$ or
$\mu_{d}$ higher than $400-500~MeV$ where other difermion
condensates might arise (see for instance ref.
\cite{Alford:2002rz}).

Let us now consider the Lagrangian of the NJL model with two
flavors $u,d$ with the same mass $m$ but different chemical
potentials $\mu_{u}$ and $\mu_{d}$
\begin{eqnarray}
{\cal {L}}&=& {\cal {L}}_{0}+{\cal {L}}_{m}+{\cal {L}}_{\mu}+{\cal {L}}_{int}\nonumber\\
&=&{\bar{\Psi}} i{\hat{\partial}}\Psi ~-~ m~{\bar{\Psi}}\Psi~+~
\Psi^{\dagger}~A~\Psi~+~{G\over 2}\sum_{a=0}^{3}\left[\left(
{\bar{\Psi}}\tau_{a}\Psi
\right)^{2}+\left({\bar{\Psi}}i\gamma_{5}\tau_{a}\Psi\right)^{2}
\right] \label{eq:njlagr}
\end{eqnarray}
where $\Psi=\left(
\begin{array}{c} u\\ d
\end{array} \right)$, $~~A=\left(
\begin{array}{c} \mu_u\\0
\end{array} \begin{array}{c} 0\\\mu_d
\end{array}\right)$ is the matrix of chemical potentials and\break
$~~\tau_{a}$, $a=0,1,2,3$, is the set of the three Pauli matrices plus the identity.\\
We note that we can express ${\cal {L}}_{\mu}$ either by using the
variables $\mu_{u},\mu_{d}$ or the two combinations
$\mu_{B}=\displaystyle{{\mu_{u}+\mu_{d}\over 2}}$ and
$\mu_{I}=\displaystyle{{\mu_{u}-\mu_{d}\over 2}}$, which couple to
the baryon charge density and to the third component of isospin
respectively
\begin{equation}
{\cal {L}}_{\mu}=\mu_{B}~\Psi^{\dagger}\Psi
~+~\mu_{I}~\Psi^{\dagger}\tau_{3}\Psi
\label{eq:lmuiso}
\end{equation}
To study whether a pion condensate shows up, we need to calculate
the effective potential. This is obtained by using the standard
technique to introduce bosonic (collective) variables through the
Hubbard-Stratonovich transformation and by integrating out the
fermion fields in the generating functional. However, the
effective potential that we have considered is not directly
obtained from the Lagrangian in Eq. (\ref{eq:njlagr}). To mimic
asymptotic freedom we want to include a form factor as in ref.
\cite{Alford:1997} and we thus follow the same procedure as in
refs.  \cite{Berges:1998rc,Toublan:2003tt}. The result is a
one-loop effective potential which generalizes that of the theory
described by the Lagrangian in Eq. (\ref{eq:njlagr}), and which
reduces to it in the limit of a constant form factor $F({\vec
p})=1$.

\begin{equation}
\label{Poteff} V=\frac{\Lambda^2}{8G}
(\chi_u^2+\chi_d^2+2\rho^2)+V_{\mbox{log}}
\end{equation}

\begin{eqnarray} \label{Vlog}
V_{\mbox{log}}=-\mbox{Tr log} \left(
\begin{array}{cc}
h_u & -F^2(\vec{p})~\Lambda~\rho~\gamma_5\\
F^2(\vec{p})~\Lambda~\rho~\gamma_5 & h_d
\end{array}
\right)
\nonumber\\
\\
h_f=(i\omega_n+\mu_f)\gamma_0~-~\vec{p}\cdot\vec{\gamma}~-~
\left(m~+~F^2(\vec{p})~\Lambda~\chi_f\right)\nonumber
\end{eqnarray}
where $\omega_{n}$ are the Matsubara frequencies and the
dimensionless fields $\chi_{f}$ and $\rho$ are connected to the
condensates by the following relations
\begin{eqnarray} \label{eq:fields}
\chi_f = - ~2G~{\langle{\bar{\Psi}}_f\Psi_f\rangle\over \Lambda}\nonumber\\
\\
\rho = - ~G~{\langle\bar{u}\gamma_5d-\bar{d}\gamma_5u\rangle\over
\Lambda}\nonumber
\end{eqnarray}
and are variationally determined at the absolute minimum of the
effective potential. In the previous equations, $\Lambda$ is a
mass scale appearing in the form factor $F({\bf
p}^2)=\displaystyle{{\Lambda^2\over \Lambda^2+{\bf p}^2}}$
\cite{Alford:1997}.\\
It is worth noting that the one-loop effective potential in Eq.
(\ref{Vlog}) has the same expression of the one derived in ref.
\cite{Barducci:2003un} within $ladder$-QCD. Therein, multiplying
the scalar and pseudoscalar fields, there was a test function
guessed from the study of the one-loop Schwinger-Dyson equation
for the quark self-energy, in place of $F^2(\vec{p})$ in Eq.
(\ref{Vlog}). The only difference is that $F^2$ depends on the
three-momentum whereas the quoted test function depends on the
four-momentum and that the two asymptotic behaviors are different
($\sim 1/p^2$ in the test function of ref. \cite{Barducci:2003un}
and $\sim 1/{\vec p}~^4$ in $F^2$ of Eq. (\ref{Vlog})). Otherwise
the two effective potentials would be identical. This observation
also explains the reason why we have adopted the NJL model instead
of $ladder$-QCD to generalize the analysis of ref.
\cite{Barducci:2003un} at high isospin chemical potentials as the
numerical analysis is much simpler in this
case.\\

To fix the free parameters of the model, which are $\Lambda$, the
average current quarks mass $m=(m_{u}+m_{d})/2$ and the coupling
$G=g/\Lambda^2$, we work at zero temperature and quark densities.
We first choose the mass scale $\Lambda$ within the range $\Lambda
\sim 500-600~ MeV$. Then we determine the strength of the coupling
$g$ and the mass parameter $m$ by requiring a light quark
condensate of the order $\langle {\bar \Psi}_{f}\Psi_{f}\rangle
\simeq -(200~ MeV)^3$ and a pion mass $m_{\pi}\simeq 140~MeV$ (the
latter evaluated through the curvature of the effective potential
in the direction of the pion field and having fixed $f_{\pi}$ at
its experimental value \cite{Barducci:1988gn}).


The output parameters are the following
\begin{equation}
\Lambda=580~MeV;~~~~~~~~~~g=7;~~~~~~~~~~~m=4.5~MeV
 \label{eq:parameters}
\end{equation}
With these values we obtain a condensate $\langle{\bar
\Psi}_{f}\Psi_{f}\rangle= -(172~MeV)^3$ and a constituent quark
mass $M_{f}=428~MeV$ (defined as in
\cite{Kunihiro,Berges:1998rc}). The critical isospin chemical
potential at zero temperature turns out to be $\mu_I^C=89~MeV$.
The discrepancy of about $25\%$ (we recall that the expected value
of $\mu_I^C$ would be $m_{\pi}/2\simeq 70~MeV$) is due to the
approximate fit procedure. Actually, in our previous work based on
$ladder$-QCD \cite{Barducci:2003un}, where $f_{\pi}$ was
consistently calculated within the model,
we got $\mu_I^C=m_{\pi}/2$.\\

\section{Phase diagram for chiral symmetry breaking and pion condensation}
\label{sec:physics}

In order to discuss the structure of the phase diagram, it is
worth summarizing the symmetries of the Lagrangian density in Eq.
(\ref{eq:njlagr}). Both ${\cal {L}}_{0}$ and ${\cal {L}}_{int}$
are $SU_{L}(2)\otimes SU_{R}(2)\otimes U_{V}^{B}(1)\otimes
U^{B}_{A}(1)$ invariant. The symmetry is reduced by the mass term
${\cal {L}}_{m}$ to $SU_{V}(2)\otimes U_{V}^{B}(1)$ and further
reduced from the term ${\cal {L}}_{\mu}$ which selects a direction
in the isospin space, as is evident from Eq. (\ref{eq:lmuiso}),
unless $\mu_{u}=\mu_{d}$ and thus $\mu_{I}=0$. The remaining
symmetry can be expressed either as $U_{V}^{u}(1)\otimes
U_{V}^{d}(1)$ or $U_{V}^{B}(1)\otimes U_{V}^{I}(1)$, depending on
the basis of the fields that we are choosing.\\
The baryon number symmetry $U_{V}^{B}(1)$ is dynamically
respected, whereas a non vanishing v.e.v. of the $\rho$ field
defined in Eq. (\ref{eq:fields}) may appear, which dynamically
breaks $U_{V}^{I}(1)$. This implies the appearance of a Goldstone
mode, which is either the charged $\pi^{+}$ or $\pi^{-}$ at the
threshold, depending on the sign of $\mu_{I}$,
whereas the other two pions are massive modes.\\
As far as the scalar condensates $\chi_{u},\chi_{d}$ (see Eq.
(\ref{eq:fields})) are concerned, they do not break any symmetry.
However, since the mass term is small, their value is almost
entirely due to the approximate spontaneous breaking of chiral
symmetry. Consequently we distinguish regions where the dynamical
effect is relevant, from regions where the scalar condensates are
of order $\sim m/\Lambda$, namely where only the effect of the
explicit breaking of chiral
symmetry survives.\\
The determination of the various phases has been performed
numerically by minimizing the one-loop effective potential. We
start by showing the results in the $(\mu_{u},\mu_{d})$ plane, for
fixed values of the temperature. Different regions are labelled,
as in ref. \cite{Klein:2003fy}, by the symbol of the field which
acquires a non vanishing v.e.v. due to dynamical effects, whereas
the other fields are vanishing ($\rho$), or of the order $\sim
m/\Lambda$
($\chi_{u}$ and/or $\chi_{d}$).\\
Solid lines refer to discontinuous transitions and dashed lines to
continuous ones. However, we recall that strictly speaking only
the lines surrounding regions with a non vanishing field $\rho$
refer to genuine phase transitions, associated with the breaking
and restoration of the $U_{V}^{I}(1)$ symmetry.

\subsection{Zero temperature}

\begin{figure}[htbp]
\begin{center}
\includegraphics[width=10cm]{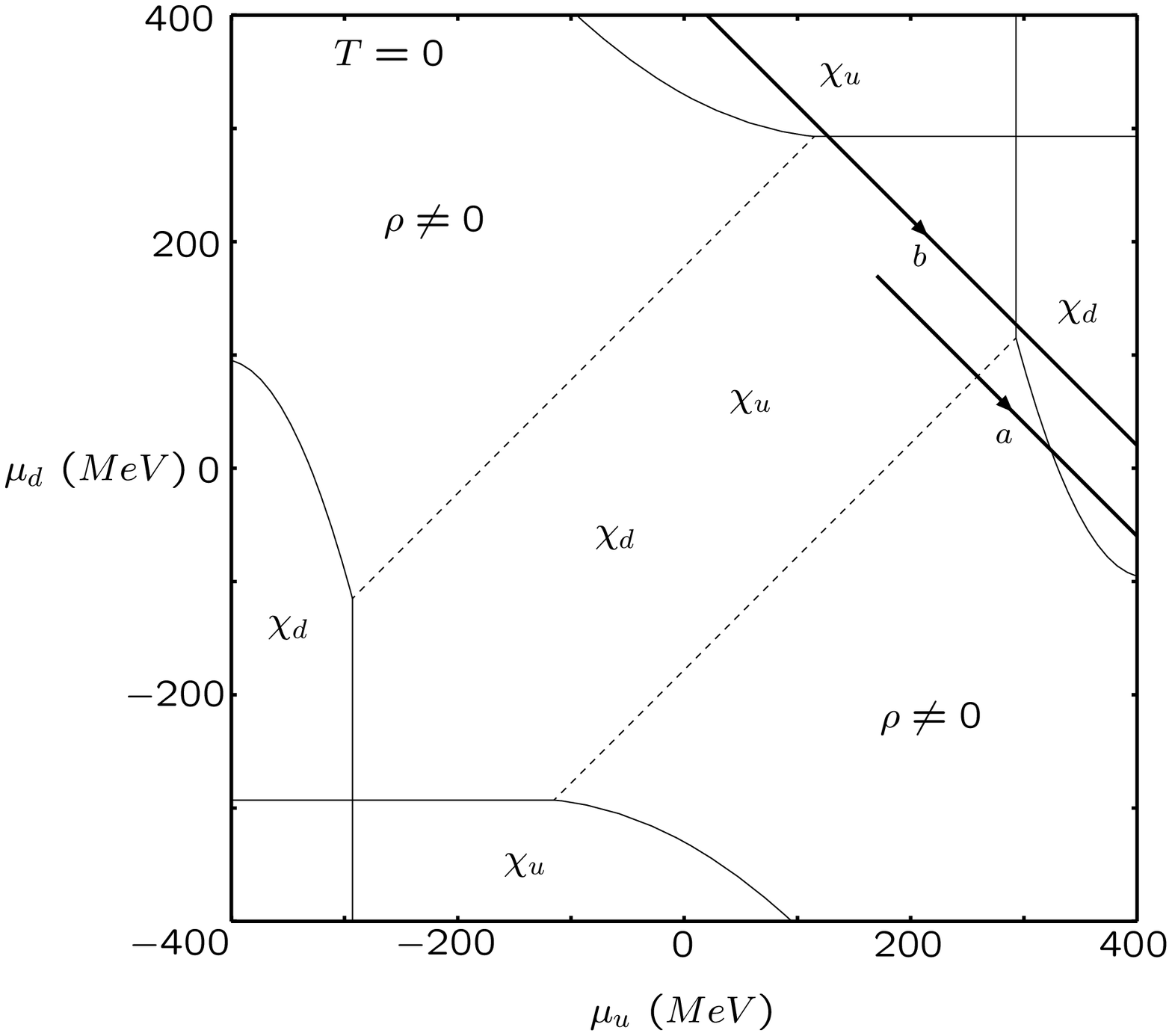}
\end{center}
\caption{\it Phase diagram for chiral symmetry restoration in the
plane $(\mu_{u},\mu_{d})$ of quark chemical potentials, at $T=0$.
Different regions are specified by the non vanishing of a given
condensate, whereas the others are vanishing ($\rho$) or order
$\sim m/\Lambda$ ($\chi_{u}$ and $\chi_{d}$). Dashed lines are for
the continuous vanishing of $\rho$ or for cross-over phase
transitions for $\chi_{u}$ or $\chi_{d}$, whereas solid lines are
for discontinuous behaviors. The solid lines $a$ and $b$ refer to
specific paths at fixed values of $\mu_{B}$, with
$\mu_{B}=170~MeV$ (line $a$) relative to Fig. \ref{fig:mufetta100}
and $\mu_{B}=210~MeV$ (line $b$) relative to Fig.
\ref{fig:mufetta250}.} \label{fig:totdiamuftz1}
\end{figure}

\begin{figure}[htbp]
\begin{center}
\includegraphics[width=10cm]{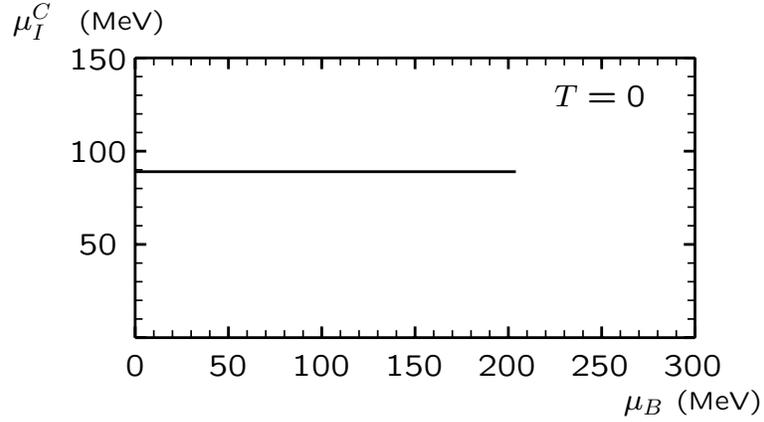}
\end{center}
\caption{\it Critical value of the isospin chemical potential,
beyond which a pseudoscalar condensate forms, vs. baryon chemical
potential, at zero temperature. $\mu_{B}=204~MeV$ is the highest
allowed value for pion condensation to occur. The path followed in
the phase diagram of Fig. \ref{fig:totdiamuftz1} is along the
upper-half of the dashed line in the lower half-plane.}
\label{fig:critisovsmub}
\end{figure}

\begin{figure}[htbp]
\begin{center}
\includegraphics[width=10cm]{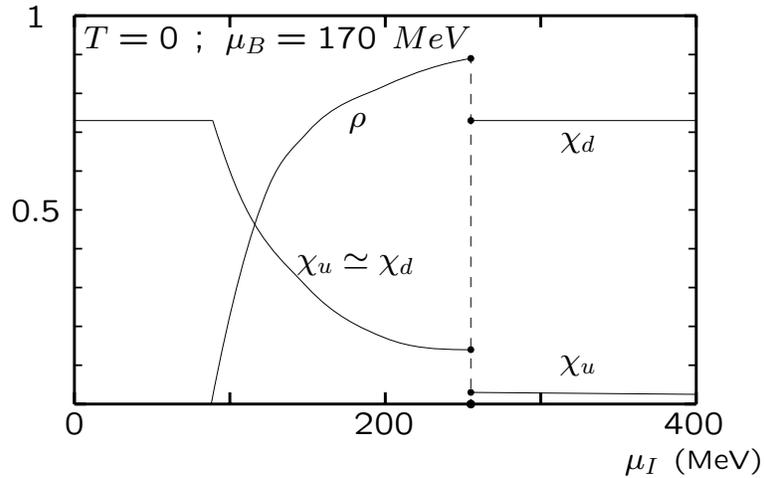}
\end{center}\caption{\it Scalar and pseudoscalar
condensates vs. $\mu_I$, for $\mu_B=170~MeV$ and $T=0$. The path
followed in the phase diagram of Fig. \ref{fig:totdiamuftz1} is
that of the solid line $a$.} \label{fig:mufetta100}
\end{figure}

\begin{figure}[htbp]
\begin{center}
\includegraphics[width=10cm]{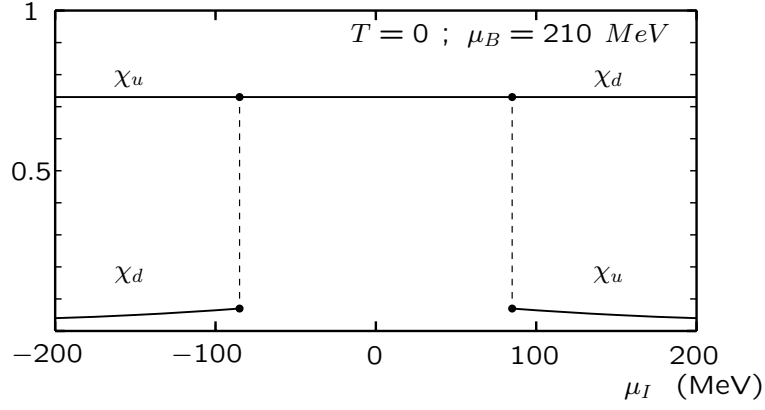}
\end{center}
\caption{\it
Scalar condensates vs. $\mu_{I}$ for $\mu_{B}=210~MeV$ and $T=0$.
The pseudoscalar condensate is zero. The path followed in the
phase diagram of Fig. \ref{fig:totdiamuftz1} is that of the solid
line $b$.} \label{fig:mufetta250}
\end{figure}

\begin{figure}[htbp]
\begin{center}
\includegraphics[width=10cm]{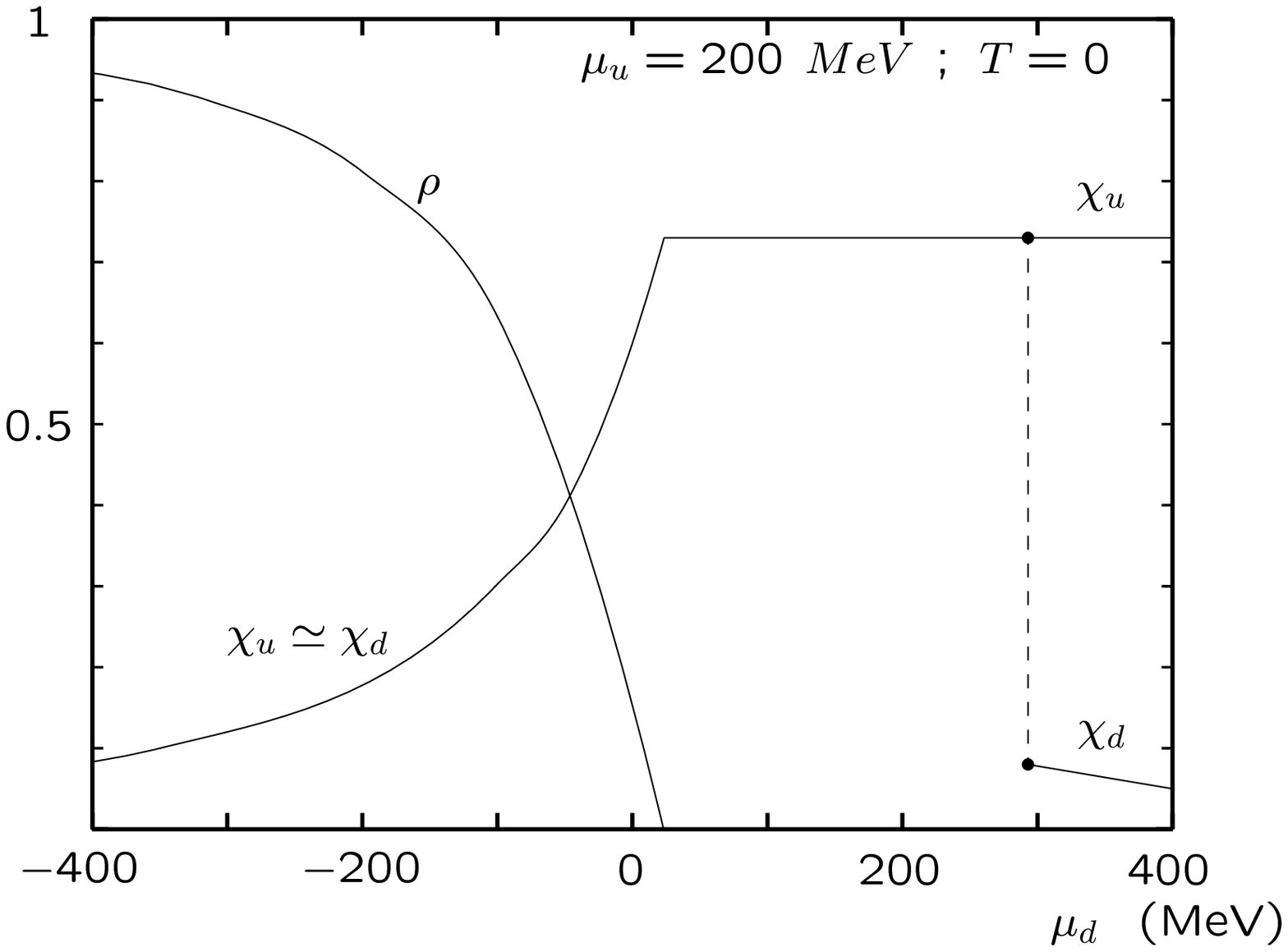}
\end{center}
\caption{\it Scalar and pseudoscalar condensates vs. $\mu_d$, for
$\mu_u=200~MeV$, $T=0$ (see Fig. \ref{fig:totdiamuftz1}).}
\label{fig:tfetta2}
\end{figure}

In Fig. \ref{fig:totdiamuftz1} we show the phase diagram in the
$(\mu_{u},\mu_{d})$ plane at zero temperature. Let us start from
the vacuum at $T=\mu_{u}=\mu_{d}=0$, at the center of the picture.
Here, in the chiral limit, the pions are the Goldstone bosons
associated with the spontaneous breaking of $SU(2)_L\otimes
SU(2)_R$. We have chosen these variables in order to compare the
structure of the phase diagram with that obtained in ref.
\cite{Klein:2003fy}. However, if we want to recover known results
given in terms of the baryon chemical potential, we have to move
from the center along the diagonal at $\mu_u=\mu_d$ and thus at
$\mu_{I}=0$, and increase the absolute value of $\mu_{B}$. At
$\displaystyle{{|\mu_{u}+\mu_{d}|\over 2}=|\mu_{B}|}=293~MeV$ we
meet the approximate restoration of chiral symmetry due to the
sudden jump of the condensates of the two (degenerate) quarks to
values of order $\sim m/\Lambda$, which is a discontinuous
transition. The same thing happens by moving along lines parallel
to the main diagonal in the region labelled by
$\chi_{u},~\chi_{d}$, enclosed between the two dashed lines at
$\displaystyle{{|\mu_{u}-\mu_{d}|\over 2}}=|\mu_I|= 89~MeV$ (see
also Fig. \ref{fig:critisovsmub} where it is shown that the
critical value of $\mu_{I}$ at $T=0$ is independent on $\mu_{B}$)
and by varying $|\mu_{B}|$. The regions in the top-right and
bottom-left corners of Fig. \ref{fig:totdiamuftz1} thus have the
$U_{V}^{u}(1)\otimes U_{V}^{d}(1)$ symmetry of ${\cal {L}}$ with
$\rho=0$ and $\chi_{u}$,$\chi_{d}$ of order $\sim m/\Lambda$.\\
By moving from the center along the diagonal at $\mu_{u}=-\mu_{d}$
(and thus $\mu_{B}=0$) or parallel to it, when crossing one of the
two dashed lines at $|\mu_{I}|=89~MeV$ (we have already discussed
the origin of the discrepancy between this value and half of the
pion mass in the model), the absolute minimum of the effective
potential starts to rotate along the $\rho$ direction. We thus
have a continuous breaking of $U_{V}^{I}(1)$ and a second order
phase transition with one Goldstone mode which is, right along the
dashed line, either the $\pi^{+}$ (in the upper part of the
diagram) or the $\pi^{-}$ (in the lower part). In the chiral limit
these two dashed lines merge together in coincidence with the
diagonal at $\mu_{I}=0$ as the pion becomes massless in this limit
and the rotation is sudden, giving first order phase transitions
for pion condensation. In this case there are two Goldstone bosons
associated with the spontaneous breaking of two $U(1)$ symmetry
groups
($U_{A}^{B}(1)\otimes U_{V}^{I}(1)$) \cite{Kogut:2002zg}.\\
Coming back to the massive case, and still with reference to Fig.
\ref{fig:totdiamuftz1}, we conclude that by considering
$|\mu_{B}|$ not too large and by growing $|\mu_{I}|$, we find a
second order phase transition with the rotation of the scalar
condensates into the pseudoscalar, namely we are faced with pion
condensation in a relatively simple picture. A difference with
ref. \cite{Klein:2003fy} is that we do not find the vanishing of
$\rho$ for values of $|\mu_{I}|$ high with respect to the pion
mass, but still sufficiently low to avoid considering
superconductive phases (actually, for very low $|\mu_{B}|$ this
transition would occur in the present model for $|\mu_{I}|\sim 1~GeV$).\\
To explore the possibility of multiple phase transitions and thus
of a richer phenomenology, we need to grow $|\mu_{B}|$ as for
instance we do in the case described in Fig. \ref{fig:mufetta100}
where we follow the path of the solid line $a$ in Fig.
\ref{fig:totdiamuftz1} at $\mu_{B}=170~MeV$ for growing
$\mu_{I}\geq 0$. The fields $\chi_{u}$ and $\chi_{d}$ are almost
degenerate, both in the region of the approximate dynamical
breaking of chiral symmetry (below $\mu_{I}=89~MeV$) and in the
region of spontaneous breaking of $U_{V}^{I}(1)$, where they
rotate into the $\rho$ field. Then, when the line $a$ in Fig.
\ref{fig:totdiamuftz1} crosses the solid line surrounding the
region labelled by $\chi_{d}$, we see that $\rho$ suddenly jumps
to zero with the restoration of the $U_{V}^{I}(1)$ group and that
$\chi_{u}$ and $\chi_{d}$ split. Actually the latter suddenly
acquires a value due to the dynamical breaking of chiral symmetry
whereas $\chi_{u}$ undergoes a further decrease and remains of
order $\sim m/\Lambda$.\\
In Fig. \ref{fig:mufetta250} we plot the behavior of the scalar
condensates $\chi_{u}$ and $\chi_{d}$ vs. $\mu_{I}$ at
$\mu_{B}=210~MeV$, namely by following the path described by the
solid line $b$ in Fig. \ref{fig:totdiamuftz1}. We see that we
never cross the region with $\rho\neq 0$ and that we simply pass
from a region where the dynamical effect of the breaking of chiral
symmetry is entirely due to a large value of $\chi_{u}$ and
$\chi_{d}$ of order $\sim m/\Lambda$ (at large negative $\mu_{I}$
and small $\mu_{u}$), to a region where this effect manifests
itself with a large value of $\chi_{d}$ and $\chi_{u}$ of order
$\sim m/\Lambda$ (at large positive $\mu_{I}$ and small
$\mu_{d}$). The region in between has almost degenerate and both
large $\chi_{u}$ and $\chi_{d}$. Pion condensation does not occur
for this value of $\mu_B$
(see also Fig. \ref{fig:critisovsmub}).\\
Finally, in Fig. \ref{fig:tfetta2}, we plot the behavior of the
condensates at fixed $\mu_u=200~MeV$ vs. $\mu_d$ (see again Fig.
\ref{fig:totdiamuftz1}). The rotation of the pion condensate into
the scalar ones occurs when the vertical line at $\mu_u=200~MeV$
meets the dashed line at $\mu_I=89~MeV$, which happens for $\mu_d$
of few $MeV$. Then, when $\mu_d$ has sufficiently increased,
$\chi_d$ falls to a small value of order $\sim m/\Lambda$ with a
discontinuous transition, whereas $\chi_u$ remains constant at its
large value.

\subsection{Finite temperature}

\begin{figure}[htbp]
\begin{center}
\includegraphics[width=10cm]{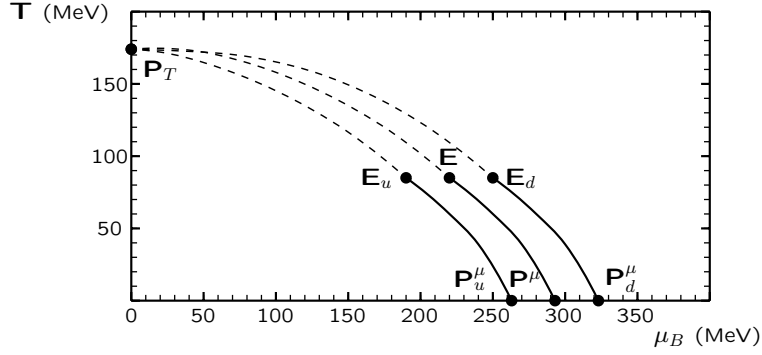}
\end{center}
\caption{\it{Phase diagram for chiral symmetry in the
$(\mu_{B},T)$ plane for zero or small isospin chemical potential
$\mu_{I}$. For $\mu_{I}=0$ (central line), the cross-over
transition line starts from the point $P_{T}=(0,174)$ and ends at
the point $E=(220,85)$. The line between $E$ and the point
$P^{\mu}=(293,0)$ is the line for the first order transition with
discontinuities in the $\langle{\bar u}u\rangle$ and $\langle{\bar
d}d\rangle$ condensates. For $\mu_{I}=30~{\rm MeV}$ (side lines),
the two cross-over transition lines start from the point
$P_{T}=(0,174)$ and end at the points $E_{u}=(190,85)$ and
$E_{d}=(250,85)$. The lines between $E_{u}$ and the point
$P^{\mu}_{u}=(263,0)$ and between $E_{d}$ and the point
$P^{\mu}_{d}=(323,0)$ are the lines for the first order
transitions with discontinuities in the $\langle{\bar u}u\rangle$
and $\langle{\bar d}d\rangle$ condensates respectively.}}
\label{fig:diafaisopic}
\end{figure}

\begin{figure}[htbp]
\begin{center}
\includegraphics[width=10cm]{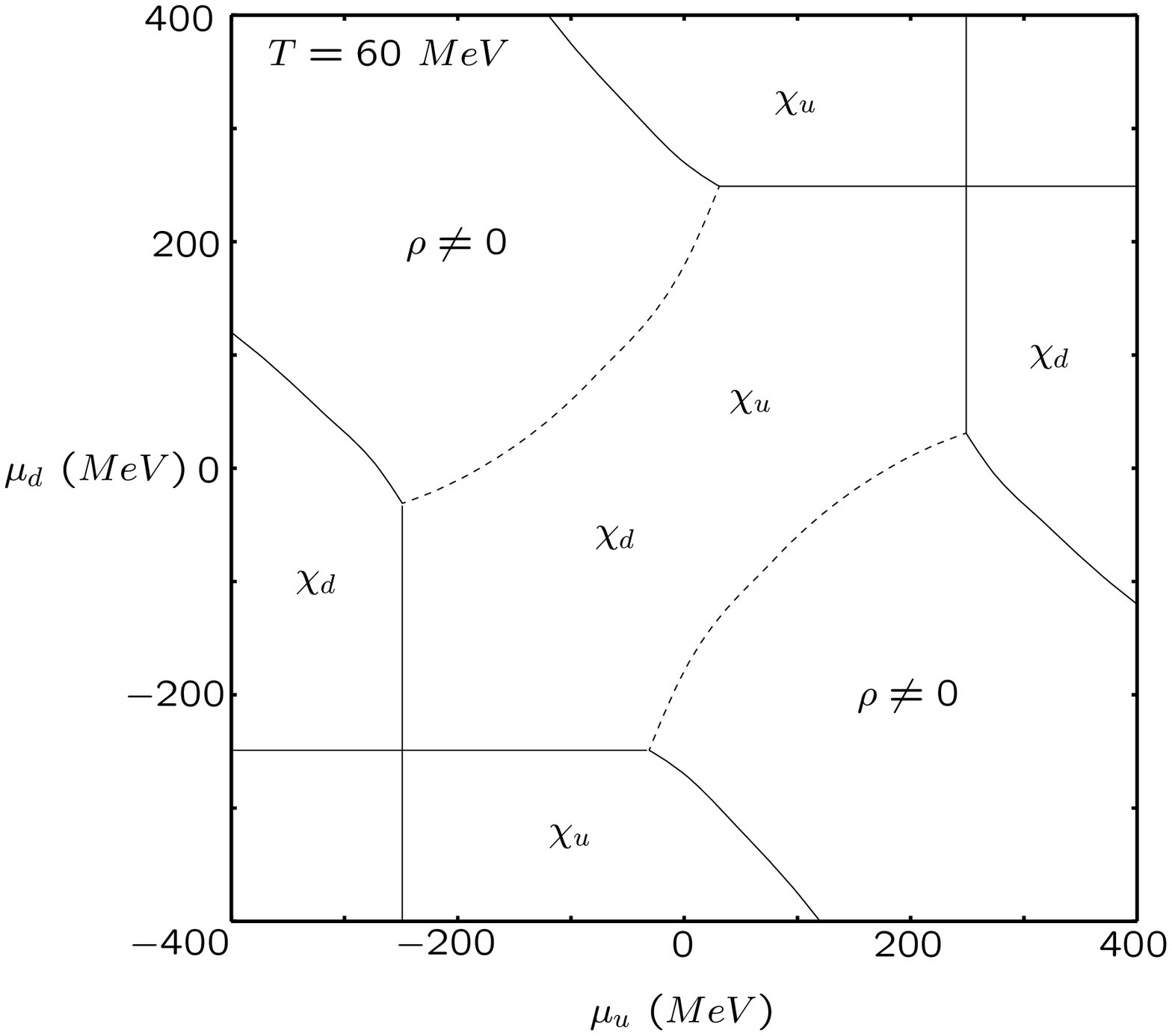}
\end{center}
\caption{\it Phase diagram for chiral symmetry restoration in the
plane $(\mu_{u},\mu_{d})$ of quark chemical potentials, at
$T=60~MeV$, which is below the temperature of the critical ending
point (see Fig. \ref{fig:diafaisopic}). Different regions are
specified by the non vanishing of a given condensate, whereas the
others are vanishing ($\rho$) or $\sim m/\Lambda$ ($\chi_{u}$ and
$\chi_{d}$). Dashed lines are lines for the continuous vanishing
of $\rho$ or for cross-over phase transitions for $\chi_{u}$ or
$\chi_{d}$, whereas solid lines are for discontinuous behaviors.}
\label{fig:totdiamuftz2}
\end{figure}

\begin{figure}[htbp]
\begin{center}
\includegraphics[width=10cm]{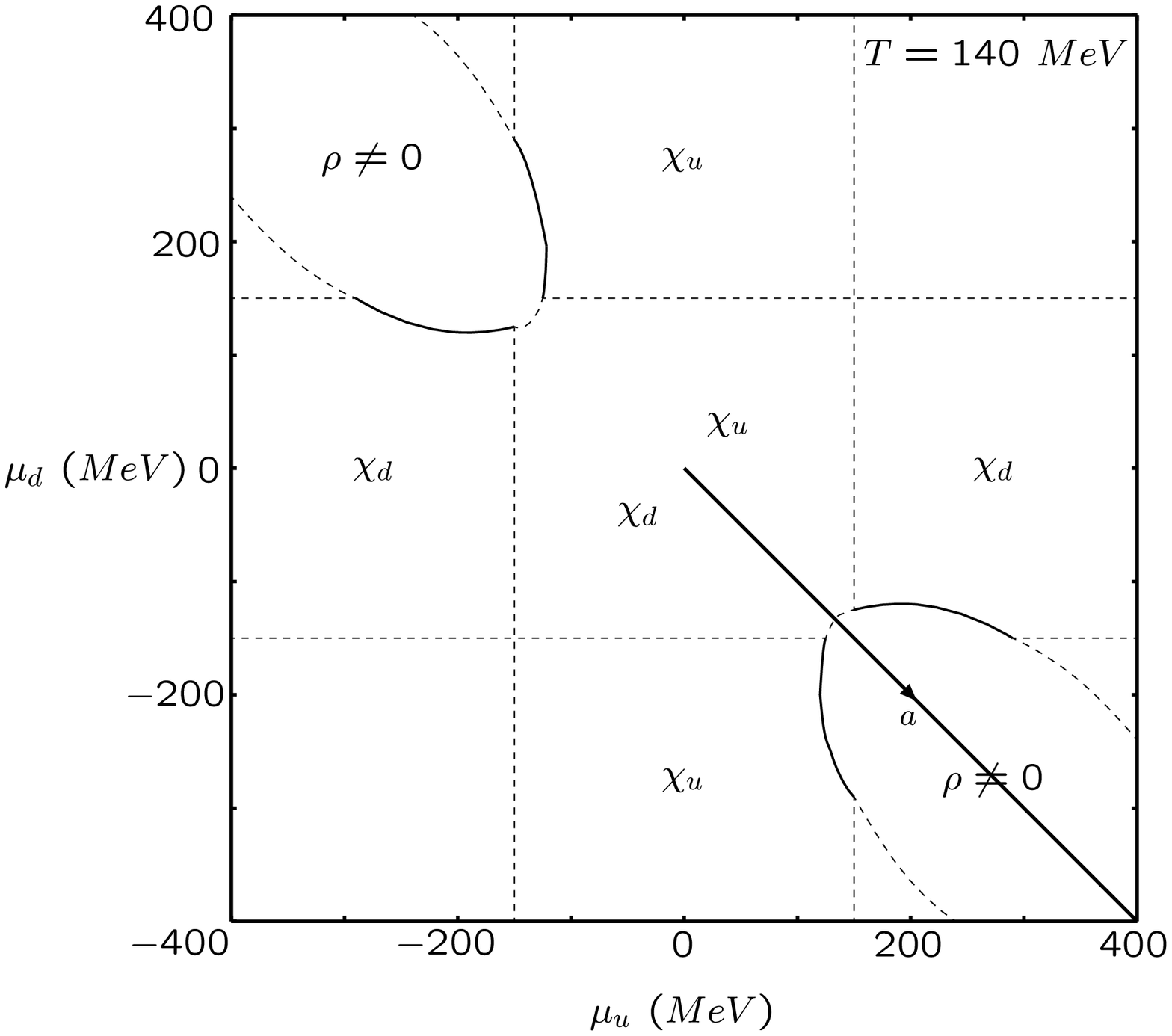}
\end{center}
\caption{\it  Phase diagram for chiral symmetry restoration in the
plane $(\mu_{u},\mu_{d})$ of quark chemical potentials, at
$T=140~MeV$, which is above the temperature of the critical ending
point (see Fig. \ref{fig:diafaisopic}). Different regions are
specified by the non vanishing of a given condensate, whereas the
others are vanishing ($\rho$) or $\sim m/\Lambda$ ($\chi_{u}$ and
$\chi_{d}$). Dashed lines are lines for the continuous vanishing
of $\rho$ or for cross-over phase transitions for $\chi_{u}$ or
$\chi_{d}$, whereas solid lines are for discontinuous behaviors.
The solid line $a$ refer to the path at $\mu_{B}=0$ vs.
$\mu_{I}\geq 0$ followed in Fig. \ref{fig:tfetta150}.}
\label{fig:totdiamuftz3}
\end{figure}

\begin{figure}[htbp]
\begin{center}
\includegraphics[width=10cm]{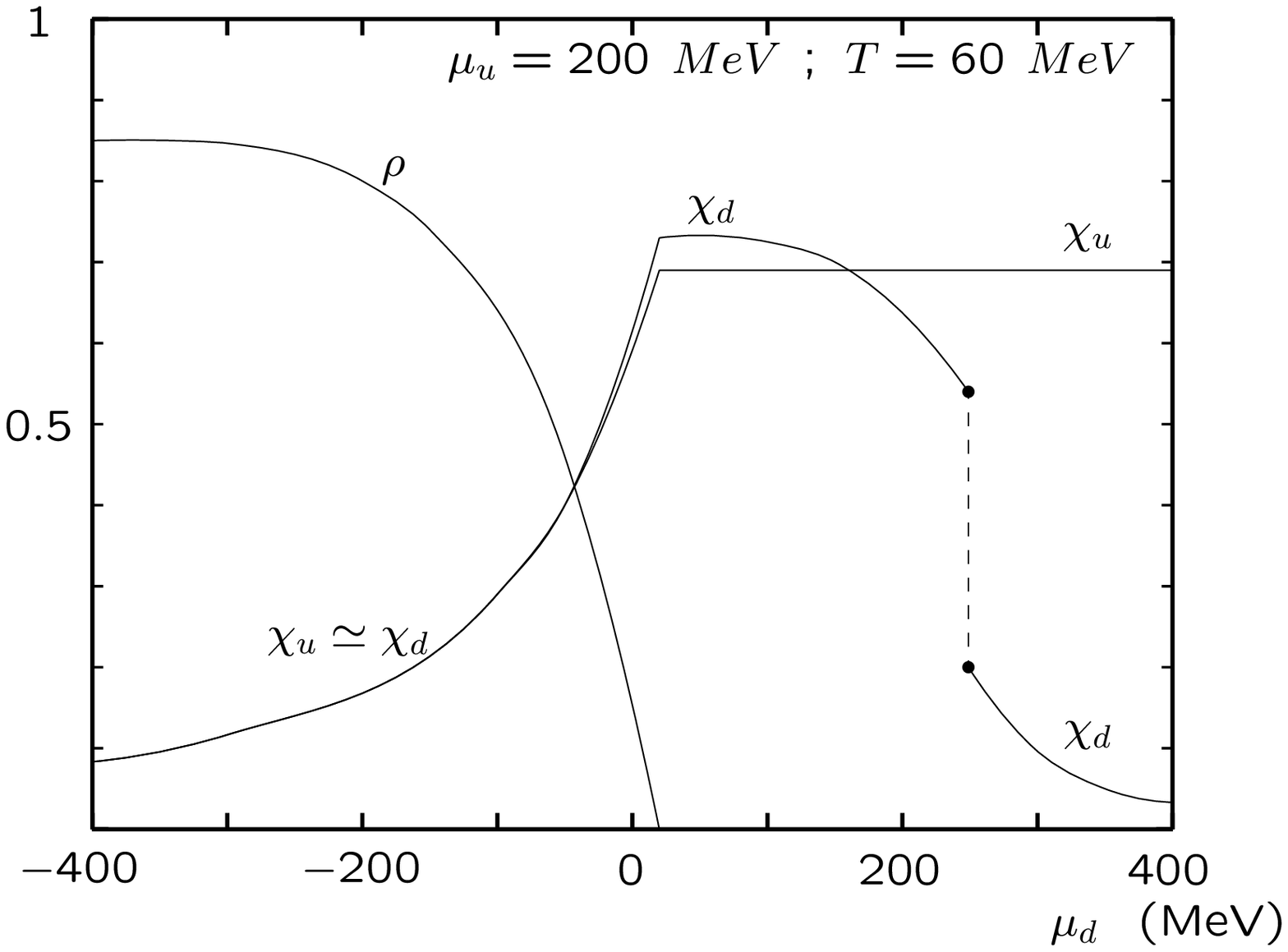}
\end{center}
\caption{\it Scalar and pseudoscalar condensates vs. $\mu_d$, for
$\mu_u=200~MeV,~T=60~MeV$ (see Fig. \ref{fig:totdiamuftz2}).}
\label{fig:tfetta5}
\end{figure}

\begin{figure}[htbp]
\begin{center}
\includegraphics[width=10cm]{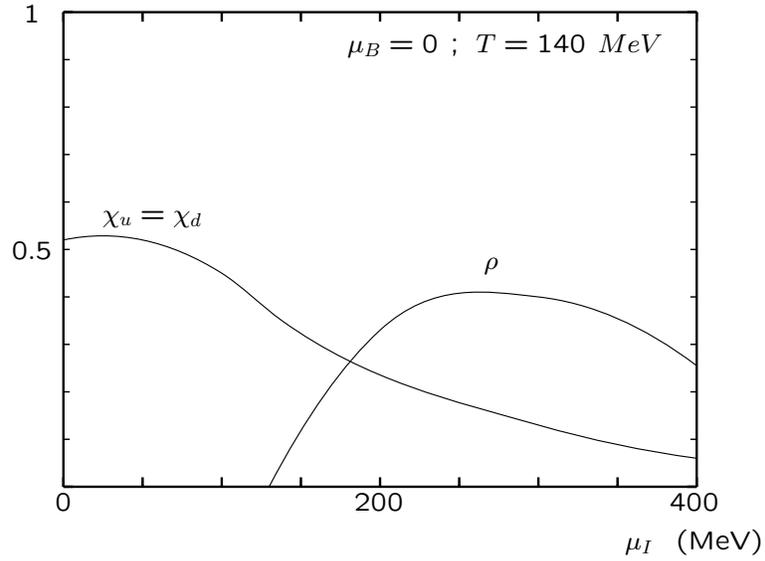}
\caption{\it Scalar and pseudoscalar condensates vs. $\mu_I$, for
$\mu_B=0$ and $T=140~MeV$. The figure is obtained following the
path $a$ in the phase diagram of Fig. \ref{fig:totdiamuftz3}.}
\label{fig:tfetta150}
\end{center}
\end{figure}

\begin{figure}[htbp]
\begin{center}
\includegraphics[width=10cm]{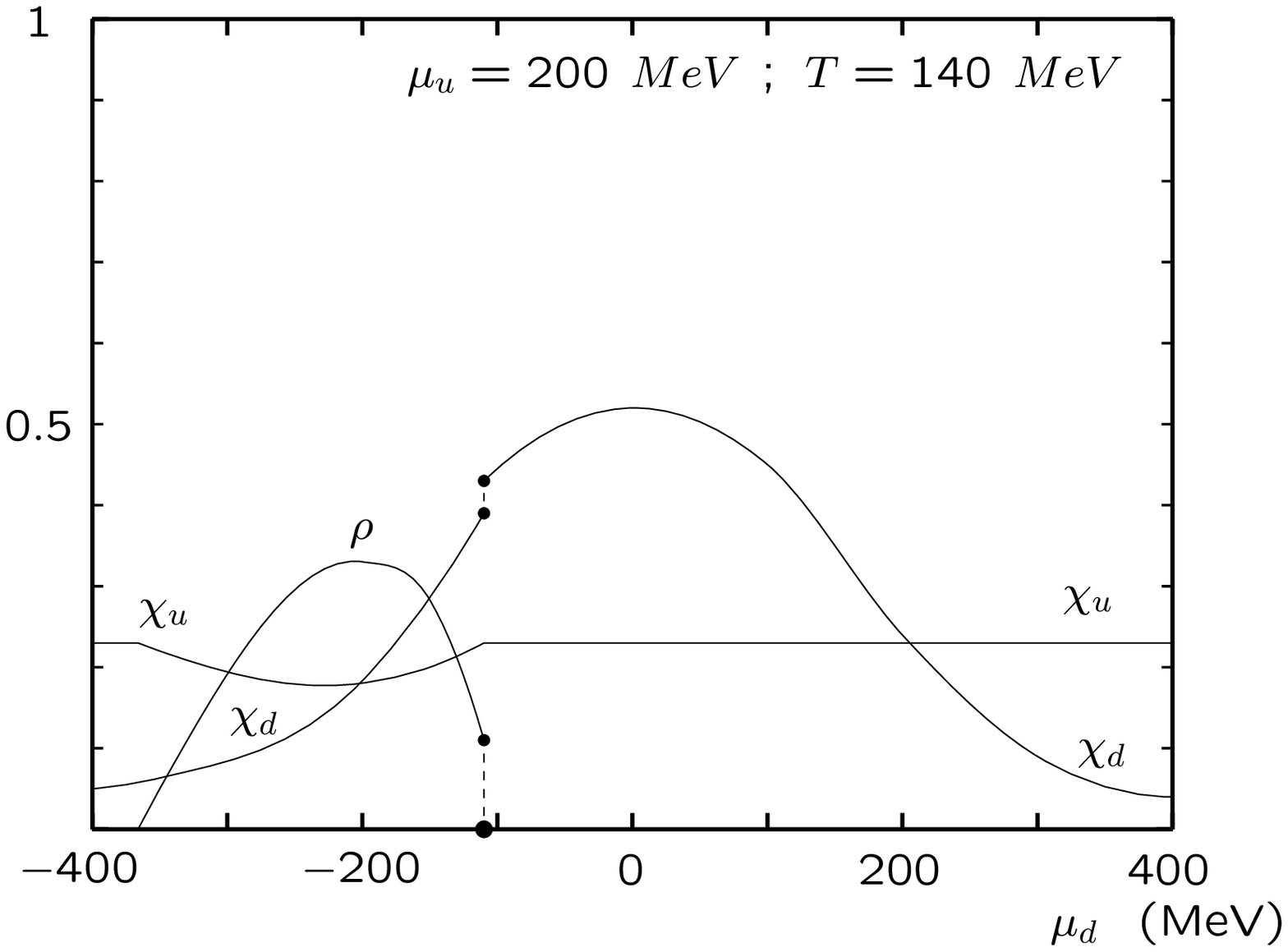}
\end{center}
\caption{\it Scalar and pseudoscalar condensates vs. $\mu_d$, for
$\mu_u=200~MeV,~T=140~MeV$ (see Fig. \ref{fig:totdiamuftz3}). }
\label{fig:tfetta33}
\end{figure}

\begin{figure}[htbp]
\begin{center}
\includegraphics[width=10cm]{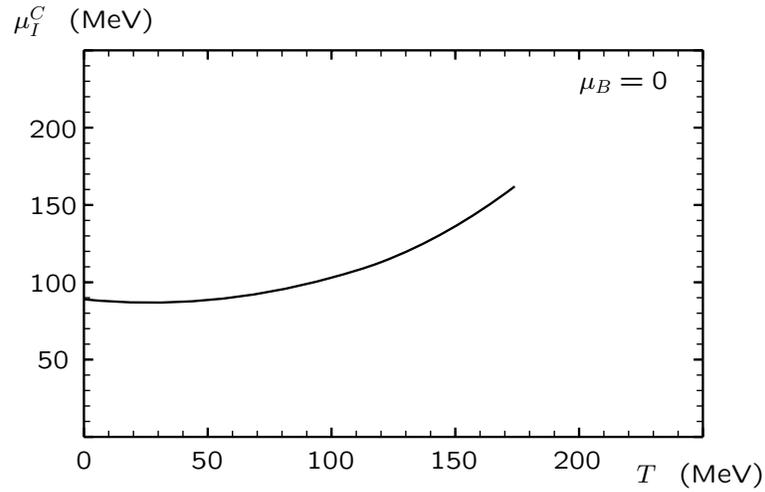}
\end{center}
\caption{\it Critical value of the isospin chemical potential,
beyond which a pseudoscalar condensate forms, vs. temperature, at
zero baryon chemical potential.} \label{fig:critisovst}
\end{figure}

\begin{figure}[htbp]
\begin{center}
\includegraphics[width=10cm]{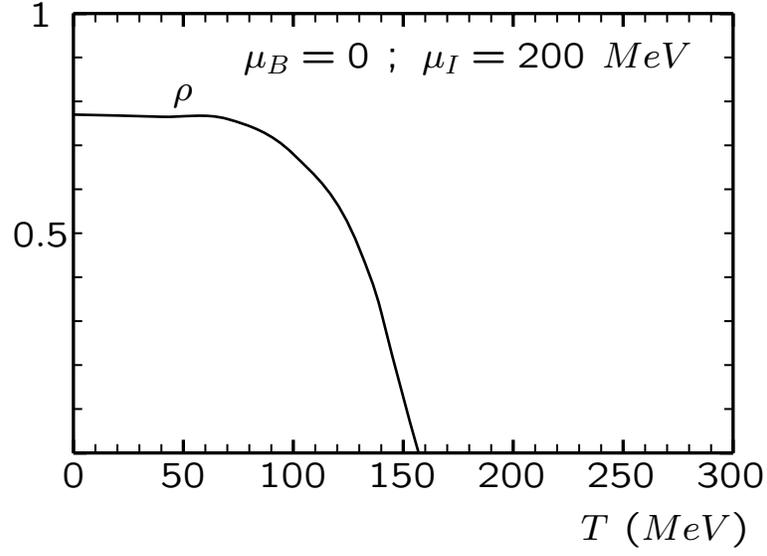}
\end{center}
\caption{\it Pion condensates vs. $T$, for $\mu_B=$ and
$\mu_{I}=200~MeV$. The scalar condensates $\chi_{u}$ and
$\chi_{d}$ are of order $\sim m/\Lambda$.}
\label{fig:ifetta200vst}
\end{figure}

The evolution of the phase diagram for growing temperatures is
easily understood as far as the regions with $\rho=0$ are
concerned. Actually, in this case the effective potential at the
minimum is the sum of two independent terms, one for each flavor,
and the results are straightforwardly given through the analysis
of chiral symmetry breaking and restoration for a single flavor at
finite temperature and chemical potential (see for instance refs.
\cite{Barducci:1990wi,Barducci:1993bh}). In Fig.
\ref{fig:diafaisopic} we show the phase diagram at zero, or small
isospin chemical potential (see also refs.
\cite{Toublan:2003tt,Barducci:2003un}). From this picture we see
that moving along any of the critical lines of chiral symmetry
restoration at fixed $\mu_{I}$, the critical value of the baryon
chemical potential $\mu_{B}$ decreases for growing temperatures.
Furthermore, for temperatures below that of the critical ending
point $E$,  $T<T(E)=85~MeV$, the transitions are always
discontinuous whereas they become cross-over transitions for
$T>T(E)$. Consequently the regions labelled by $\chi_{u}$ and/or
$\chi_{d}$ in Fig. \ref{fig:totdiamuftz1} shrink when growing $T$
and their rectilinear sides become lines of cross-over transitions
for $T> T(E)$ (see Fig. \ref{fig:totdiamuftz1} and Figs.
\ref{fig:totdiamuftz2}, \ref{fig:totdiamuftz3} where we plot the
phase diagrams in the $(\mu_{u},\mu_{d})$ plane at $T=60~MeV$ and
$T=140~MeV$, which is respectively below and above $T(E)$). The
new feature concerns the regions with $\rho\neq 0$, which also
reduce their size for growing $T$, whereas the order of the
transitions starts to change from first to second, beginning from
the critical points at highest $|\mu_{I}|$, until they reach the
points of the boundaries which coincide with those of the regions
labelled by $\chi_{u}$ or $\chi_{d}$ (see Fig.
\ref{fig:totdiamuftz3}). Also the length of the curves of second
order phase transitions to pion condensation at fixed values of
$|\mu_{B}|$ sensibly reduces from low temperatures to high
temperatures (see again Fig. \ref{fig:totdiamuftz1}, and Figs.
\ref{fig:totdiamuftz2}, \ref{fig:totdiamuftz3}).  A similar
behavior, for high $T$, is found in ref. \cite{Klein:2003fy}. For
$T> T(P_T)=174~MeV$, which is the cross-over temperature at zero
chemical potential (see Fig. \ref{fig:diafaisopic}), all these
regions disappear from the phase diagram, which is thus
characterized by $\rho=0$ and
$\chi_{u},\chi_{d}~\sim ~ m/\Lambda$.\\
The behavior of the scalar and pseudoscalar condensates at
$T=60~MeV$ are much similar to those at $T=0$. As an example we
plot, in Fig. \ref{fig:tfetta5}, the condensates at
$\mu_u=200~MeV$ and $T=60~MeV$ vs. $\mu_{d}$ (compare with Fig.
\ref{fig:tfetta2}). The situation is different if we consider
temperatures above $T(E)=85~MeV$. For instance, at $T=140~MeV$, we
see from Fig. \ref{fig:totdiamuftz3}, that the structure of the
phase diagram is only slightly modified with respect to the case
of two independent flavors which undergo cross-over phase
transitions at sufficiently high values of their own chemical
potentials (actually the region of pion condensation has sensibly
reduced with respect to Figs. \ref{fig:totdiamuftz1},
\ref{fig:totdiamuftz2}). The phase transition associated with the
spontaneous breaking of $U_{V}^{I}(1)$ can be both second or first
order, depending on the path followed. In Fig.
\ref{fig:totdiamuftz3}, the solid line $a$ refers to a path at
$\mu_{B}=0$ vs. $\mu_{I}\geq 0$ where the transition to pion
condensation is continuous. The behavior of the condensates
relative to this path is plotted in Fig.. \ref{fig:tfetta150}. In
Fig. \ref{fig:tfetta33} we plot the scalar and pseudoscalar
condensates
at $\mu_{u}=200~MeV$ and $T=140~MeV$ vs. $\mu_{d}$. \\
In Fig. \ref{fig:critisovst} we plot the value of the critical
isospin chemical potential $\mu_{I}^C$ beyond which a pion
condensate forms vs. temperature $T$ at zero baryon chemical
potential $\mu_{B}$. The growth of $\mu_{I}^{C}$ is easily
understood since the pion mass (which should be twice
$\mu_{I}^{C}$) is expected to grow near the critical temperature
for chiral symmetry restoration \cite{Barducci:1991rh}. On the
other hand, no phase transition to pion condensation is expected
above the cross-over temperature for chiral symmetry restoration.
Thus the line of
critical values ends at $T=174~MeV$.\\
The pion condensate is also expected to decrease for growing
temperatures: in Fig. \ref{fig:ifetta200vst} we show $\rho$ vs.
$T$ at $\mu_{B}=0$ and $\mu_{I}=200~MeV$. Similar behaviors are
obtained for fixed values of $\mu_{B}$ and $\mu_{I}$.\\

\section{Conclusions}

In this paper we have continued the study of pion condensation at
finite baryon and isospin density in the NJL model that we started
in ref. \cite{Barducci:2003un} in the case of $ladder$-QCD for
small isospin chemical potentials. The extension to higher isospin
chemical potentials confirms the structure predicted in ref.
\cite{Klein:2003fy}, where two-flavor QCD was simulated in the
context of a random matrix model. Some difference between the two
analysis is present, at low temperatures, in the region of high
isospin chemical potentials, at the boundary of the region where
color superconductivity should take place. Actually in this region
we find that pion condensation is still active, whereas in ref.
\cite{Klein:2003fy} the authors find that the pion condensate
vanishes. We have also shown the expected behavior of scalar and
pion condensates by following different paths, for growing
temperatures, both in the plane of quark chemical potentials
$(\mu_{u},\mu_{d})$ and in that of isospin and baryon chemical
potentials $(\mu_{I},\mu_{B}$). The analysis that we have
performed should also be confirmed, with only small quantitative
differences, within $ladder$-QCD.



\eject

\end{document}